\documentstyle[preprint,aps]{revtex}
\begin{document}
\preprint{OITS-563, UH-511-810-94, NHCU-HEP-94-30}
\draft
\title{CP Violation In Hyperon Decays Due To Left-Right Mixing}
\author{Darwin Chang$^{1,4}$, Xiao-Gang He$^2$ and Sandip Pakvasa$^3$}
\address{$^1$ Department of Physics, National Tsing Hua University\\
Hsinchu, 30043, Taiwan \\
\\
$^2$Institute of Theoretical Science, University of Oregon\\
Eugene, OR 97403-5203, USA\\
\\
$^3$ Department of Physics and Astronomy, University of Hawaii\\
Honolulu, Hawaii, HI 96822, USA
\\
$^4$ Institute of Physics, Academia Sinica\\
Nankang, Taipei, 115 Taiwan
}
\date{December 1994}
\maketitle
\begin{abstract}
We consider CP violation due to left-right mixing in a class of Left-Right
symmetric models and show that it leads to observable effects in hyperon
decays. For S-wave, the contribution to CP
violating asymmetry A of the polarization in hyperon and anti-hyperon
decays is proportional to $\epsilon'/\epsilon$.
While the tree level L-R operators contribution is
constrained to be less than $ 10^{-5}$, the gluon penguin operator
contribution can be as large as $10^{-4}$.  For P-wave, the
contribution is not directly related to $\epsilon'/\epsilon$. Its contribution
to A can be larger and may reach $6\times 10^{-4}$ in $\Lambda \rightarrow
p\pi^-$.  This is much larger than the value expected for A in the Standard
Model.
\end{abstract}
\newpage
Non-leptonic hyperon decays of $\Lambda$, $\Sigma$ and
$\Xi$ are interesting processes to test CP conservation
outside the neutral Kaon system\cite{pre,darwin,dhp,sm,new}.
Non-leptonic hyperon decays proceed into both S-wave (parity-violating) and
P-wave (parity-conserving) final states with amplitudes S and P, respectively.
One can write the amplitude in the rest frame of the initial baryon as
\begin{eqnarray}
Amp(B_i \rightarrow B_f\pi) =S + P\vec \sigma\cdot \vec q\;,
\end{eqnarray}
where, $B_i, B_f$ are initial and final baryons, $\vec q$ is the momentum of
pion. It is convenient to write the amplitudes
as
\begin{eqnarray}
S &=& \sum_i S_i e^{i(\phi^S_i + \delta_i^S)}\nonumber\\
P &=& \sum_i P_i e^{i(\phi^P_i + \delta_i^P)}\;
\end{eqnarray}
to explicitly separate the strong rescattering phases $\delta_i$ and the weak
CP
violating phases $\phi_i$. Here $i$ is summed over different isospin
channels. One particularly interesting observable
is the asymmetry A defined in Ref.\cite{dhp}
\begin{eqnarray}
A = {\alpha+\bar \alpha\over \alpha-\bar \alpha}\;,
\end{eqnarray}
where $\alpha = 2 Re(S^*P)/(|S|^2 +|P|^2)$, and $\bar \alpha$ is the
corresponding quantity for anti-hyperon decays. A non-zero A signals CP
violation.

Recently the E871 proposal
at Fermilab has been approved\cite{E871}. The
expected sensitivity for CP violation in $\Xi^- \rightarrow \Lambda \pi^-$ and
$\Lambda \rightarrow p\pi^-$ at E871 may reach the upper range of the
Standard Model (SM) prediction\cite{dhp,sm,new}. It is interesting to see
if with non SM sources these CP
asymmetries may be
large, and what kind of constraints the experimental data may be able to impose
on extensions of the SM.
In this paper we study these CP asymmetries due to left-right mixing
in a class of Left-Right Symmetric Models\cite{darwin}.

The gauge group of the left-right symmetric extension of the SM is
$SU(3)_C\times
SU(2)_L\times SU(2)_R\times U(1)_{B-L}$\cite{lrm,conj}. In this model there are
two charged
gauge bosons $W_L$ and $W_R$ associated with the $SU(2)_L$ and $SU(2)_R$ gauge
groups. In general there will be mixing between $W_L$ and $W_R$. We follow the
convention used in Ref.\cite{lr} The mass
eigenstates $W_{1,2}$ can be parameterized as
\begin{eqnarray}
W_L = cos\xi W_1 -sin\xi W_2\;, W_R = sin\xi W_1 +cos\xi W_2\;,
\end{eqnarray}
with the convention being that $W_1$ being the lighter of two. The charged
current interaction in the quark sector has the form
\begin{eqnarray}
L &=& -{1\over 2\sqrt{2}} \bar U\gamma_\mu [g_L cos\xi V_L(1-\gamma_5) +
g_R sin\xi V_R(1+\gamma_5)]D W^\mu_1\nonumber\\
&-&{1\over 2\sqrt{2}} \bar U\gamma_\mu [-g_L sin\xi V_L(1-\gamma_5) +
g_R cos\xi V_R(1+\gamma_5)]D W^\mu_2 + h.c.\;,
\end{eqnarray}
where $g_{L,R}$ are the $SU(2)_{L,R}$ gauge couplings,  $V_{L,R}$ are the KM
mixing matrices, and
$U = (u,c,t,...)$ and $D = (d,s,b,...)$.
The model in general has $N^2 - N + 1$ CP violating phases with or without
left-right discrete symmetry for N generations.
In models with left-right discrete symmetry, one has $g_L = g_R$.  If, in
addition, the CP symmetry is assumed to be only spontaneously broken,
it is possible to choose a phase convention such that $V_L = V_R^*$ and
therefore the number of CP violating phases is reduced to $N(N+1)/2$
\cite{lrcp}.  Note that, with or without the discrete symmetry, even in the
one generation case, the theory has one genuine CP violating phase which
is the relative phase between $V_L$ and $V_R$.
This is the phase that can be generically identified as the left-right
mixing phase.

We write the effective Hamiltonian for non-leptonic hyperon decays up to one
loop level as
\begin{eqnarray}
H_{eff} = H_{SM} + H_{R} + H_{LR}\;.
\end{eqnarray}
Here $H_{SM}$ indicates the effective Hamiltonian generated by exchanging $W_1$
at the tree and one loop level.
In the zero mixing limit, this reduces to the SM contribution. In the following
we will keep terms up to linear in $\xi$ because it is constrained to be small
\cite{mixing}. To the leading order, $H_{SM}$ is
given by\cite{desh}
\begin{eqnarray}
H_{SM} &=& {G_F\over \sqrt{2}} \{ V_{Lud}^*V_{Lus}\bar d \gamma_\mu
(1-\gamma_5)u \bar u\gamma_5(1-\gamma_5) s \nonumber\\
&+& \sum_i V_{Lid}^*V_{Lis} {\alpha_s \over 8\pi} E(x_i) \bar d
\gamma_\mu(1-\gamma_5)\lambda^a s \sum_{u,d,s} \bar q \gamma_\mu \lambda^a
q\nonumber\\
&+& \sum_i V_{Lid}^*V_{Lis}G(x_i) {g_s \over 32\pi^2}m_s \bar d
\sigma_{\mu\nu}(1+\gamma_5)\lambda^a s G^{a\mu\nu}\}\;,
\end{eqnarray}
where $x_i = m_i^2/m_{W_1}^2$, $ i=u,\;c,\;t$, $G^{a\mu\nu}$ is the gluon field
strength,
 the functions $E(x)$ and $G(x)$ are give by
\begin{eqnarray}
E(x) &=& -{2\over 3} \mbox{ln}x + {x(18-11x-x^2)\over 12(1-x)^3}
+{x^2(15-16x+4x^2)\over 6(1-x)^4}\mbox{ln}x\;,\nonumber\\
G(x) &=& -{3x^2\over 2(1-x)^4}\mbox{ln}x + {x^3-5x^2-2x\over 4(1-x)^3}\;.
\end{eqnarray}
The first term in eq.(7) is due to exchange of $W_1$ at the tree level, the
second term is the gluon penguin operator, and the third is the gluon dipole
penguin operator.

$H_R$ is due to $W_2$ exchange. It can be obtained from $H_L$ by
replacing $(1\pm\gamma_5)$ with $(1\mp\gamma_5)$,
$m_{W_1}$ by $m_{W_2}$, $G_F$ by $G_F (g_R^2m^2_{W_1}/g_L^2m_{W_2}^2)$, and
$V_L$ by $V_R$. $H_{LR}$ is due to $W_L$ and $W_R$
mixing and therefore should be proportional to the mixing matrix elements
of $ \xi V_{L(R)}V^*_{R(L)}$.  It is given by
\begin{eqnarray}
H_{LR} &=&  {G_F\over \sqrt{2}} \bar \xi
\{V^*_{Lud}V_{Rus}
\bar d \gamma^\mu (1-\gamma_5)u
\bar u \gamma_\mu (1+\gamma_5)s
+ V^*_{Rud} V_{Lus}
\bar d \gamma^\mu (1+\gamma_5)u
\bar u \gamma_\mu (1-\gamma_5)s
\nonumber\\
&+& \sum_i \tilde G(x_i) {g_s\over 16 \pi^2} m_i
G^{a\mu\nu} \bar d \sigma_{\mu\nu}\lambda^a
[V_{Rid}^*V_{Lis} (1-\gamma_5) + V_{Lid}^*V_{Ris}(1+\gamma_5)]s\}
\end{eqnarray}
where $\bar \xi = \xi g_R/g_L$, and
\begin{eqnarray}
\tilde G(x) = -{3x\over 2(1-x)^3}\mbox{ln}x - {4+x+x^2\over 4(1-x)^2}\;.
\end{eqnarray}

Detailed studies of $H_{SM}$ contribution to CP violation in hyperon with QCD
corrections have been carried out in Ref.\cite{dhp,sm,new}.  It was shown that
the asymmetry
$A(\Lambda)$ is in the range $-(1\sim 5) \times 10^{-5}$, and $A(\Xi)$ is in
the range $-(1\sim 10)\times 10^{-5}$ if CP violation is due to the phase in
$V_L$. In these calculations, the strong rescattering phases in
Ref.\cite{lphase} for $\Lambda$ decays were used.  For $A(\Xi)$ decays,the
S-wave rescattering phase
$\delta_2 = -18.7^0$\cite{xphase1} and the P-wave rescattering phase
$\delta_{12} = -2.7^0$\cite{xphase2}
were used. There have been new calculations for $\delta_2$ and $\delta_{12}$
using chiral perturbation theory.  The rescattering phases were
found to be smaller with $\delta_2 = 0.2$ and $\delta_{12} = -1.7^0$ from
low lying intermediate states\cite{nphase,nphase1}.  The asymmetry
$A(\Xi)$
is scaled down by a factor of 10. In our later calculation we will use the
strong rescattering phases from Ref.\cite{nphase,nphase1}.

In Left-Right Symmetric Models with N
generations of quarks, one can choose the phase convention such that only
$(N-1)(N-2)/2$ phases appear in $V_L$, while the other $N(N+1)/2$
phases in $V_R$.
It may well be the case that the source for CP violation
is purely from the phases in $V_R$ in this convention.
If $V_R$ is approximately equal to $V_L$, then the contribution to CP
violation
in hyperon decay from $H_R$ is suppressed by a factor of $m_{W_1}^2/m_{W_2}^2$
which is small. In general, $V_R$ can be very different from $V_L$. In that
case there may
be larger
contribution to CP violation in hyperon decays.
In Ref.\cite{dhp} a special case, the ``isoconjugate''  Left-Right
Model\cite{conj} where CP violation
is due to phases in $V_R$, there is no left-right mixing ($\xi = 0$), and
$|V_{Lij}| = |V_{Rij}|$, was discussed.
In this case, the leading $\Delta S =1$ Hamiltonian has the form
\begin{eqnarray}
H_{eff} = {G_F\over \sqrt{2}} \mbox{sin}\theta_c\mbox{cos}\theta_c
(O_{LL} + r e^{i\beta}O_{RR})\;,
\end{eqnarray}
where $r = m_{W_1}^2/m_{W_2}^2$, $O_{LL}$ and $O_{RR}$ are identical operators,
except that $O_{LL}$ is
a product of two left-handed currents whereas $O_{RR}$ has two right-handed
currents. Because of this structure one finds the parity-nonconserving (S-wave)
processes have an identical phase factor $1+ir\beta$, while all
parity-conserving (P-wave) ones have phase factor $1-ir\beta$. We have
\begin{eqnarray}
\phi^S_i = r\beta\;,\;\;\;\; \phi^P_i = - r\beta\;.
\end{eqnarray}
The phase $r\beta$ is determined by CP violation in neutral kaon mixing to be
about $4.4\times 10^{-5}$. Using this value for
$r\beta$,
the asymmetries
$A(\Lambda)$ and $A(\Xi)$ were predicted to be $-10^{-5}$ and $2.5\times
10^{-6}$, respectively\cite{dhp}.

In this letter we concentrate on the effect from
$H_{LR}$ which only contribute when left-right mixing is non-zero. The
contribution to S-wave from tree level left-right mixing operators has been
considered before\cite{darwin}. Here we will carry out a full calculation
including P-wave
contribution and contributions from left-right penguin operator. We will
include QCD corrections to this effective Hamiltonian. It is convenient to use
the following operator basis
\begin{eqnarray}
O^{LR}_+ &=& \bar d \gamma_\mu (1-\gamma_5)u\bar u \gamma^\mu (1+\gamma_5)s
+{2\over 3} \bar d (1+\gamma_5) s \bar u (1-\gamma_5) u\;,\nonumber\\
O^{RL}_+ &=& \bar d \gamma_\mu (1+\gamma_5)u\bar u \gamma^\mu (1-\gamma_5)s
+{2\over 3} \bar d (1-\gamma_5) s \bar u (1+\gamma_5) u\;,\nonumber\\
O^{LR}_-&=&{2\over 3} \bar d (1+\gamma_5) s \bar u (1-\gamma_5) u\;,\nonumber\\
O^{RL}_-&=&{2\over 3} \bar d (1-\gamma_5) s \bar u (1+\gamma_5) u\;.
\end{eqnarray}
We have\cite{lr}
\begin{eqnarray}
H_{LR} &=& {G_F\over \sqrt{2}}\bar \xi\{
V^*_{Lud}V_{Rus}(O_+^{LR}\eta_+ - O_-^{LR}\eta_-)
+V^*_{Rud}V_{Lus}(O_+^{RL}\eta_+ - O_-^{RL}\eta_-)\nonumber\\
&+&\sum_i \tilde G(x_i) {g_s\over 16 \pi^2} m_i\eta_g
G^{a\mu\nu} \bar d \sigma_{\mu\nu}\lambda^a
[V_{Rid}^*V_{Lis} (1-\gamma_5) + V_{Lid}^*V_{Ris}(1+\gamma_5)]s\}\;,
\end{eqnarray}
where $\eta_+ =
(\alpha_s(1GeV)/\alpha_s(m_{m_c}))^{-3/27}(\alpha_s(m_c)/\alpha_s(m_{m_b}))^{-3/25}(\alpha_s(m_b)/\alpha_s(m_{W_1}))^{-3/23}$, $\eta_- = \eta_+^{-8}$, and
$\eta_g =\eta_+^{14/3}$. In our numerical evaluation, we will use $\Lambda_4 =
0.25$ GeV for the QCD $\Lambda$ parameter.

We will first consider the operators $O_{\pm}$.
The operators $O_\pm$ contain both $\Delta I = 1/2$ and $3/2$ components. They
will contribute to the I = 1/2 and 3/2 amplitudes $a_1$ and $a_3$ in $\Lambda$
decays. In the
factorization approximation, we obtain
\begin{eqnarray}
&a_{1}&(\Lambda \rightarrow p\pi^-) = -{G_F\over \sqrt{2}}\bar \xi\nonumber\\
&\times& \{(V^*_{Lud}V_{Rus} - V_{Rud}^*V_{Lus}){2\over 27}
[8\eta_+ + \eta_-(1 + {15m_\pi^2\over 2(m_u+m_d)(m_s-m_u)})]V\nonumber\\
&+&(V^*_{Lud}V_{Rus} + V_{Rud}^*V_{Lus}){2\over 27}
[8\eta_+ + \eta_-(1 + {3m_\pi^2\over 2(m_u+m_d)(m_s-m_u)})]P \}\;,\nonumber\\
&a_{3}&(\Lambda \rightarrow p\pi^-) = {G_F\over \sqrt{2}}\bar \xi\nonumber\\
&\times&\{(V^*_{Lud}V_{Rus} - V_{Rud}^*V_{Lus}){1\over 27}
[8\eta_+ + \eta_-(1 -{6m_\pi^2\over (m_u+m_d)(m_s-m_u)})]V\nonumber\\
&+&(V^*_{Lud}V_{Rus} + V_{Rud}^*V_{Lus}){1\over 27}
[8\eta_+ - \eta_-(1 - {6m_\pi^2\over 2(m_u+m_d)(m_s-m_u)})]P\}\;,
\end{eqnarray}
where $V = \sqrt{2}F_\pi(m_\Lambda - m_p)g_{p\Lambda} \bar \psi_p
\psi_\Lambda$, and
$P = -(2F_\pi F_K m_\pi^2/m_K^2)g_{Kp\Lambda} \bar \psi_p\gamma_5\psi_\Lambda$.
Here $g_{p\Lambda} = -\sqrt{3/2}$, $g_{Kp\Lambda} = -g_{\pi
NN}(3-2\alpha)/\sqrt{3}$ with $g_{\pi NN} = 13.26$ and $\alpha = 0.64$ from
experimental data. $F_\pi = 93$ MeV and $F_K = 1.3 F_\pi$.

Using the experimental values for the CP conserving amplitudes\cite{hyperon},
we obtain the CP violating phases
\begin{eqnarray}
\phi^s_{1} =   -12.3 \xi^u_-  \;,\;\;\;\;
\phi^p_{1} &=& -0.33 \xi^u_+  \nonumber\\
\phi^s_{3} =   -230 \xi^u_-   \;,\;\;\;\;
\phi^p_{3} &=&  24.6 \xi^u_+
\end{eqnarray}
where
$\xi^i_\pm = \bar \xi  Im(V_{Lid}^*V_{Ris} \pm V_{Rid}^*V_{Lis})$.
Note that the phases in $I=3/2$ amplitudes are much larger than the ones in
$I=1/2$ amplitudes. We will need to use the full expression for the asymmetry.
We indicate the contribution to $A$ from $O_{\pm}$ by $A_W$.  We find
\begin{eqnarray}
A_W(\Lambda) &=& 1.73 \{\xi^u_-  - 0.06 \xi^u_+ \}\;.
\end{eqnarray}

The CP violating decay amplitudes for $\Xi^- \rightarrow \Lambda \pi^-$ can be
obtained in the same way. We can obtain these amplitudes by replacing $V$ and
$P$
in eq.(15) by $V'= \sqrt{2}F_\pi(m_\Xi - m_\Lambda)g_{\Xi\Lambda} \bar
\psi_\Lambda \psi_\Xi$, and
$P' = -(2F_\pi F_K m_\pi^2/m_K^2)g_{K\Lambda\Xi} \bar
\psi_\Lambda\gamma_5\psi_\Xi$. Here $g_{\Xi\Lambda} = \sqrt{3/2}$, and
$g_{K\Lambda\Xi} = -g_{\pi NN}(4\alpha-3)/\sqrt{3}$, respectively. We obtain
\begin{eqnarray}
A_W(\Xi) &=& 0.28 \{ \xi^u_- +  0.09 \xi^u_+ \}\;.
\end{eqnarray}
We note that the asymmetry $A$ in $\Xi$ decay is smaller than that in $\Lambda$
decay.

The gluon dipole penguin is a pure $\Delta I = 1/2$ operator. To estimate its
contributions
to CP violation in hyperon decays, we use MIT bag model to calculate the
hadronic matrix element $<\pi^- p|\bar d \sigma_{\mu\nu}
(1,\;\gamma_5)\lambda^aG^{a\mu\nu}s|\Lambda>$ following Ref.\cite{dhp}. We use
the CP violating phases calculated in Ref.\cite{dhp} for Weinberg model and
normalize them to the
coefficients in our model. We define the normalization factor $R$ such that
the CP violating phases $\phi_i$ is equal to $R$ times the phases given by
Donoghue, He and Pakvasa in eq.(4.8)
and (4.9) of Ref.\cite{dhp}.  We obtain the normalization factor for S-wave as
\begin{eqnarray}
R_s = {G_F\over \sqrt{2}} \sum_i
\xi^i_- {\tilde G(x_i)\over 16\pi^2} {m_i\over m_s}
\eta_g g_s \tilde A {1\over M_{K\pi}}\;,
\end{eqnarray}
where $\tilde A = 0.4$ GeV$^3$, $m_s = 0.28$ GeV is the mass used in the MIT
bag model calculation\cite{dh1}, and the parameter $M_{K\pi}$ is equal to
$5.8\times 10^{-11}$ GeV\cite{dh}. The normalization factor for P-wave is
obtained by replacing $\xi^i_-$ by $ - \xi^i_+$.
We finally have
\begin{eqnarray}
\phi^s_{1} = - 1.65 \sum_i
\xi^i_- \tilde G(x_i) {m_i\over GeV}
\;,\;\;\;\;
\phi^p_{1} = + 1.95 \sum_i
\xi^i_+ \tilde G(x_i) {m_i\over GeV}\;.
\end{eqnarray}
The gluon dipole penguin contribution $A_G$ to the asymmetry $A$ is given by
\begin{eqnarray}
A_G(\Lambda) = 0.21 \sum_i
\tilde G(x_i){m_i\over GeV}
\{ \xi^i_- + 1.17 \xi^i_+ \}\;.
\end{eqnarray}
Similarly, for $\Xi^-\rightarrow \Lambda \pi^-$, we obtain
\begin{eqnarray}
A_G(\Xi) = -2.8\times 10^{-2} \sum_i
\tilde G(x_i){m_i\over GeV}
\{ \xi^i_- - 0.5 \xi^i_+ \}\;.
\end{eqnarray}

The mixing angle is constrained to be less than $4\times
10^{-3}$\cite{mixing}. If
the magnitude of $V_{Rud(s)}$ is about the same as $V_{Lud(s)}$, that
is $Im(V^*_{L,Rud}V_{R,Lus})\approx 0.2\mbox{sin}(phase)$, the asymmetry
$A_W(\Lambda)$ can be as large as $10^{-3}$. However, this will
not be the case. The same operators will also contribute to
$\epsilon'/\epsilon$. In Ref.\cite{cplr} it was shown that $\epsilon'/\epsilon
\sim 1.25\times 10^{3} \xi^u_-$.
The S-wave
contribution to $A$ is proportional to $\epsilon'/\epsilon$. Using
$|\epsilon'/\epsilon| < 3 \times 10^{-3}$\cite{pd}, the S-wave contribution to
$A$ is constrained to be less than $4\times 10^{-6}$. In order to satisfy the
constraint on
$\epsilon'/\epsilon$, $\xi^u_-$
has to be
less than $2\times 10^{-6}$. If
$\xi^u_+$
is the same order as
$\xi^u_-$,
then $A$ is less than
$10^{-5}$.
If $\xi^u_-$
is small by cancellation,
the P-wave contribution is approximately given by
\begin{eqnarray}
A_W(\Lambda) &=& -0.2\bar\xi Im(V_{Rud}^*V_{Lus})\;.
\end{eqnarray}
$Im(V_{Rud}^*V_{Lus})$ is bounded to be less than 0.22. We see that
$A_W(\Lambda)$ can be as large as $10^{-4}$ if
$Im(V_{Rud}^*V_{Lus}) >0.1$, and such a possibility is not ruled out.

The magnitude for $A_G(\Lambda)$ is also constrained. The gluon dipole
operator also
contributes to $\epsilon'/\epsilon$\cite{dh} which is approximately given by
$\epsilon'/\epsilon \sim 5 \sum_i
\tilde G(x_i) m_i/GeV \xi^u_- $\cite{new,dipole}. The S-wave
contribution to $A_G(\Lambda)$ is
constrained to be less than $1.3\times 10^{-4}$. If the S-wave and the P-wave
contributions are about the same order of magnitudes, one would have
$A_G(\Lambda) < 2.6\times 10^{-4}$.
However the P-wave contribution may be larger.
Assuming
 $Im(V_{Lid}^*V_{Ris})=Im(V_{Rid}^*V_{Lis})$, with $m_t = 174$ GeV and $m_c =
1.35$ GeV, we have
\begin{eqnarray}
A_G(\Lambda) &=& -48\bar \xi Im(V_{Ltd}^*V_{Rts})
-0.7\bar \xi Im(V_{Lcd}^*V_{Rcs})\;.
\end{eqnarray}
If $|V_{Ltd}| \approx |V_{Lub}| = 0.003$, and the magnitude of $V_{Rts}$ is
approximately the same as $|V_{Lcb}|$, then the first term is constrained to be
less than $2.5 \times 10^{-5}$. However the second term may still have large
contribution. $Im(V_{Lcd}^*V_{Rcs})$ can be as large as 0.22. If this factor
takes its maximal value, $A_G(\Lambda)$ can be as large as $6\times 10^{-4}$
which will be probed by the E871 experiment. The asymmetry $A$ in $\Xi$ decay
is always constrained to be less than $10^{-4}$.

In conclusion, we have studied
 CP violation in hyperon decays due to left-right mixing in Left-Right
Symmetric Model. The S-wave contribution to CP violating asymmetry A
is constrained from $\epsilon'/\epsilon$. The contribution from the tree level
L-R operators  is
constrained to be less than $ 10^{-5}$, while the gluon penguin operator
contribution can be as large as $10^{-4}$. The P-wave
contribution is not directly related to $\epsilon'/\epsilon$. Its contribution
to A can be larger and may reach $6\times 10^{-4}$ in $\Lambda \rightarrow
p\pi^-$. This is much larger than the SM prediction.
 The
asymmetry in $\Xi$ decay is about one order of magnitude smaller.

\acknowledgments
This work was supported in part by the US Department of Energy Grant No.
DE-FG06-85ER40224 and DE-AN03-76SF00235 and by National Science
Council(Taiwan) Grant No. 8219481.  DC wishes to thank the High Energy
Physics Group
at the University of Hawaii
for hospitality while this work was in progress.

\end{document}